\documentclass[seceq]{ptptex}

\usepackage{graphicx}
\usepackage{wrapft}

\newcommand{\e}{\mathrm{e}}
\newcommand{\ep}{\epsilon}
\newcommand{\vev}[1]{\left\langle #1 \right\rangle}
\newcommand{\SL}{S_\Lambda}
\newcommand{\SF}{S_{F,\Lambda}}
\newcommand{\SI}{S_{I,\Lambda}}
\newcommand{\lb}{\left\lbrace}
\newcommand{\rb}{\right\rbrace}
\newcommand{\Ld}[1]{\frac{\overrightarrow{\delta}}{\delta #1}}
\newcommand{\Rd}[1]{\frac{\overleftarrow{\delta}}{\delta #1}}
\newcommand{\K}[1]{K \left( #1/\Lambda\right)}
\newcommand{\nn}{\nonumber}
\newcommand{\Tr}{\mathrm{Tr}\,}
\newcommand{\Op}{\mathcal{O}}
\newcommand{\N}{\mathcal{N}}

\title{
Phase structure of a three-diensional Yukawa model%
}

\author{
Hidenori \textsc{Sonoda}%
}

\inst{
Physics Department, Kobe University, Kobe 657-8501, Japan
}

\abst{
    We use the method of the exact renormalization group to study the
    renormalization group flows of an O(N) invariant Yukawa model in
    three dimensional Euclidean space consisting of one real scalar
    and N real spinor fields.  We obtain a phase structure similar to
    that of the N-vector model with cubic anisotropy, possessing a
    region of parameters exhibiting a first order transition.  The
    particular case with one real fermion (N=1) belongs to the same
    universality class as the Wess-Zumino model with one
    supersymmetry.  For the critical exponents of the Wilson-Fisher
    type fixed points, our 1-loop approximations are generally
    consistent with the results of previous studies.
}

\begin{document}

\maketitle

\section{Introduction\label{introduction}}

The purpose of this paper is to study a model of a real scalar field
interacting with an arbitrary number of real spinor fields in three
dimensional space-time.  As is well known, the properties of a spinor
such as the number of independent components depend highly on the
dimensionality of space-time. \cite{Gliozzi:1976qd} In three
dimensional space-time, spinors transform under $SL(2,R)$, and the
real and imaginary parts do not mix.  Under dimensional reduction, a
Majorana spinor (or equivalently a Weyl spinor and its complex
conjugate) in four dimensional space-time gives a complex spinor in
three dimensions, which is a pair of real spinors.  Hence, a single
real spinor is intrinsically an object in three dimensional
space-time.  Our model with an odd number of spinors cannot be
obtained from a four dimensional model by dimensional reduction.
Therefore, we need a formalism that caters specifically to the three
dimensional space-time.

We study our model using the framework of the exact renormalization
group (ERG) upon which reviews from various viewpoints are available.
\cite{Becchi:1996an,Morris:1998da,Aoki:2000,Bagnuls:2000ae,Berges:2000ew,Pawlowski:2005xe,Gies:2006wv,Delamotte:2007pf,Igarashi:2009tj,Rosten:2010vm}
ERG is known for its successes in non-perturbative applications to
many different areas of physics such as critical phenomena, gauge
theories, lattice models, non-relativistic few-body problems, strongly
correlated electrons, and quantum gravity. \cite{ERG2010}

In this paper we are particularly interested in obtaining the phase
structure of our model.  To obtain the phase structure it is
sufficient to derive the RG flows.  To simplify this task, we restrict
our model so that it is accessible from the Gaussian fixed point.
With the $O(N)$ symmetry among $N$ real fermions, this choice leaves
only three parameters to consider.  Since we are mainly interested in
understanding the general features of the model, we do not attempt to
take advantage of the full potential of ERG.  We solve the ERG
differential equation for the Wilson action perturbatively at 1-loop
level to derive the RG flows of the parameters.

The primary finding of this paper is the phase structure similar to
that of the N-vector model with cubic
anisotropy. \cite{PhysRevB.18.1406} We find two non-trivial fixed
points (besides the fixed point of the $\phi^4$ theory).  One is the
Wilson-Fisher type with only one relevant parameter.
\cite{Wilson:1973jj} For $N=1$, this is the same as the fixed point of
the $\mathcal{N} = 1$ supersymmetric Wess-Zumino model.  For $N$ even,
this is the fixed point of the Gross-Neveu model with $N/2$ complex
fermions. \cite{Rosenstein:1988pt} The other fixed point has two
relevant directions, and its existence implies a region of the
parameter space that exhibits a first order transition.

Our work has been partially motivated by a recent work on the
$\mathcal{N} = 1$ supersymmetric Wess-Zumino model in three dimensions
\cite{Synatschke:2010ub}, where a non-trivial Wilson-Fisher type fixed
point has been discovered.  Our model includes the supersymmetric
model as a subset, and we expect to obtain the same fixed point
without imposing supersymmetry.  If $N$ is even, the model can be
rewritten for $N/2$ complex fermions, and this class of models has
been studied with ERG in Ref.~\citen{Rosa:2000ju}, where the flow equation
is solved numerically with the initial condition corresponding to the
Gross-Neveu model.

The paper is organized as follows.  In \S \ref{Yukawa model} we
introduce our model and discuss its symmetry.  In \S \ref{Wilson
  action} we define the Wilson action and its parameters.  In \S
\ref{RG equations} we derive RG flows and find fixed points.  In \S
\ref{Phase transitions} we discuss the nature of the phase transition
shown by the model.  In \S \ref{Comparison} (for $N=1$) and \S
\ref{GN} (for $N$ even) we compare our results with those of previous
studies.  Finally, in \S \ref{Conclusions} we conclude the paper with
remarks.  The extensive appendices provide technical details that make
the paper self-contained.

Throughout the paper we work in three dimensional Euclidean space, and
we use the following notation for momentum integrals:
\begin{equation}
\int_p \equiv \int \frac{d^3 p}{(2\pi)^3}
\end{equation}

\section{Yukawa model\label{Yukawa model}}

We consider a model whose classical lagrangian is given by
\begin{equation}
\mathcal{L} = \frac{1}{2} \nabla \phi \cdot \nabla \phi +
\frac{m^2}{2} \phi^2 + \frac{1}{2}  \tilde{\chi}^I
\vec{\sigma} \cdot \nabla \chi^I + \frac{g}{\sqrt{N}} \phi \frac{1}{2} 
\tilde{\chi}^I \chi^I + \frac{\lambda}{4!} \phi^4
\label{Yukawa}
\end{equation}
where $\phi$ is a real scalar, and $\chi^I\,(I=1,\cdots,N)$ are real
spinors.  (See Appendix \ref{app-spinors} for the corresponding
lagrangian in Minkowski space.)  We denote
\begin{equation}
\tilde{\chi} \equiv \chi^T \sigma_y .
\end{equation}
We adopt the Einstein convention for summation over the repeated index
$I$.

The lagrangian is invariant under the following $\mathbf{Z_2}$
transformation
\begin{equation}
\phi (x) \longrightarrow - \phi (-x),\quad
\chi^I (x) \longrightarrow i \chi^I (-x) .
\end{equation}
This invariance forbids the mass term $\tilde{\chi}^I \chi^I$.
Depending on how this discrete symmetry is realized, we expect two
phases:
\begin{enumerate}
\item $\mathbf{Z_2}$ exact --- The expectation value of the scalar
    vanishes $\vev{\phi} = 0$, and the fermions stay massless.
\item $\mathbf{Z_2}$ spontaneously broken --- $\vev{\phi} \ne 0$, and
    the fermions become massive.
\end{enumerate}
In both phases we expect the $O(N)$ symmetry among $N$ real fermions
are unbroken.

For $N=1$, we expect that the model belongs to the same universality
class as the $\mathcal{N} = 1$ Wess-Zumino model whose classical
lagrangian is given by
\begin{equation}
\mathcal{L}_{WZ} = \frac{1}{2} \left( \left(\nabla \phi\right)^2 +
    \tilde{\chi} \vec{\sigma} \cdot \nabla \chi \right) + g \phi
\frac{1}{2} \tilde{\chi} \chi + \frac{g^2}{8} \left(\phi^2 -
    v^2\right)^2
\label{Wess-Zumino}
\end{equation}
where
\begin{equation}
m^2 = - \frac{1}{2} g^2 v^2\,.
\end{equation}
The Wess-Zumino model is a subset of the Yukawa model, satisfying the
relation
\begin{equation}
\lambda = 3 g^2\,.
\end{equation}
Hence, the critical exponents of the Wess-Zumino model must be the
same as those of the Yukawa model.

\begin{figure}[t]
\begin{center}
\includegraphics{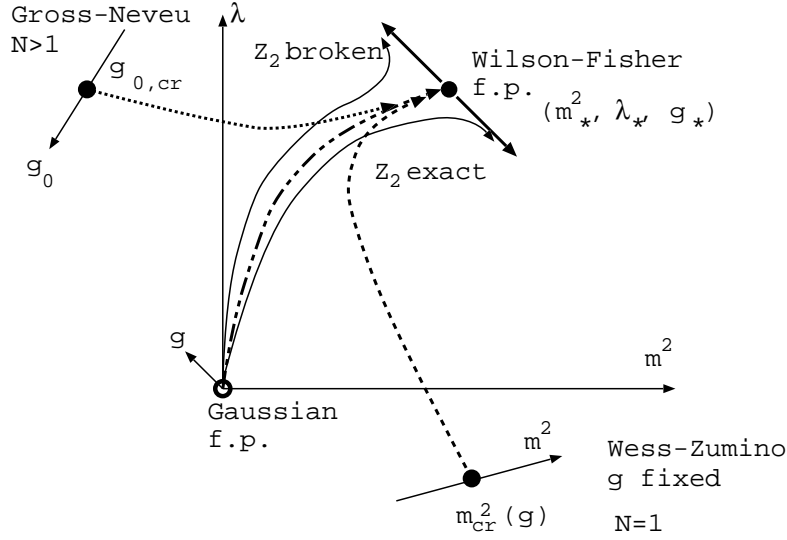}
\caption{\label{RGflows-image} Expected RG flows for the Yukawa theory
  with $N$ real fermions.  The Wess-Zumino ($N$=1) and
  Gross-Neveu models ($N > 1$) belong to the same universality class
  as the Yukawa model.}
\end{center}
\end{figure}

For $N > 1$, the Yukawa model also belongs to the same universality
class as the three dimensional Gross-Neveu model given by
\begin{equation}
\mathcal{L}_{GN} = \frac{1}{2} \tilde{\chi}^I \vec{\sigma} \cdot
\nabla \chi^I - \frac{g_0}{2 N} \left( \frac{1}{2} \tilde{\chi}^I
    \chi^I \right)^2\,.
\label{Gross-Neveu}
\end{equation}
We expect again that the critical exponents are the same as those of the
Yukawa model.  (See Fig.~\ref{RGflows-image}.)  For $N$ even, the model
contains $N/2$ complex fermions; defining complex fermions
\begin{equation}
\Psi^j \equiv \frac{1}{\sqrt{2}} \left( \chi^{2j-1} + i \chi^{2j}
\right),\quad \bar{\Psi}^j \equiv \frac{1}{\sqrt{2}} \left(
    \tilde{\chi}^{2j-1} - i \tilde{\chi}^{2j} \right)
\end{equation}
for $j = 1,\cdots, \frac{N}{2}$, we can write
\begin{equation}
\mathcal{L}_{GN} = \bar{\Psi}^j \vec{\sigma}
\cdot \nabla \Psi^j - \frac{g_0}{2 N} \left( \bar{\Psi}^j \Psi^j \right)^2\,.
\end{equation}
(See appendix \ref{app-spinors} for complex fermions.)

\section{Wilson action\label{Wilson action}}

We construct a Wilson action $\SL [\phi, \chi^I]$ that depends on a UV
cutoff $\Lambda$ in such a way that its physics is $\Lambda$
independent.\cite{Wilson:1973jj} The Wilson action is split into two
parts:
\begin{equation}
\SL [\phi, \chi^I] = \SF [\phi, \chi^I] + \SI [\phi, \chi^I]\,.
\end{equation}
The free part is given by
\begin{equation}
\SF \equiv - \int_p \frac{1}{\K{p}} \left(
\frac{1}{2} \phi (-p) (p^2 + m^2) \phi (p) + \frac{1}{2}
\tilde{\chi}^I (-p) i \vec{p} \cdot \vec{\sigma} \chi^I (p) \right)\,.
\end{equation}
The cutoff function $K(x)$ is positive, $1$ at $x^2 = 0$, and decays
fast enough for $x^2 > 1$.  To make the physics independent of
$\Lambda$, we impose that the interaction part $\SI$ satisfy the ERG
differential equation \cite{Polchinski:1983gv}
\begin{eqnarray}
&&- \Lambda \frac{\partial}{\partial \Lambda} \SI [\phi,\chi^I] =
\frac{1}{2} \int_p \frac{\Delta (p/\Lambda)}{p^2 + m^2} \lb
\frac{\delta \SI}{\delta \phi (-p)} \frac{\delta \SI}{\delta \phi (p)}
+ \frac{\delta^2 \SI}{\delta \phi (-p) \delta \phi (p)} \rb\\
&&\quad- \frac{1}{2} \int_p \frac{\Delta (p/\Lambda)}{p^2} 
\Tr (-i) \vec{p} \cdot \vec{\sigma} \lb
\Ld{\tilde{\chi}^I (-p)} \SI \cdot \SI \Rd{\chi^I (p)} 
+ \Ld{\tilde{\chi}^I (-p)} \SI \Rd{\chi^I (p)} \rb\,,\nn
\label{ERGdiff}
\end{eqnarray}
where
\begin{equation}
\label{functions}
\Delta (q) \equiv - 2 q^2 \frac{d}{dq^2} K(q)\,.
\end{equation}

To determine $\SI$ uniquely, we must introduce two additional
conditions \cite{Pernici:1998tp,Sonoda:2006ai}:
\begin{enumerate}
\item \textbf{UV renormalizability} --- We impose that the theory becomes the
    free massless theory at short distances.  We demand the following
    asymptotic conditions:
\begin{eqnarray}
    \SI &\stackrel{\Lambda \to \infty}{\longrightarrow}&
    - \int d^3 x\, \Big(
        z_{\phi, UV} \frac{1}{2} (\nabla \phi )^2 + m^2_{UV}
        \frac{1}{2} \phi^2 + z_{\chi, UV} \frac{1}{2} \tilde{\chi}^I
        \vec{\sigma} \cdot \nabla \chi^I \nn\\
&&\quad        + \frac{1}{\sqrt{N}} g_{UV} \phi \frac{1}{2}
\tilde{\chi}^I \chi^I + \lambda_{UV} \frac{1}{4!} \phi^4 \Big)
\label{UV}
\end{eqnarray}
where the terms with higher powers of fields vanish, and all the
coefficients are constant except for $m^2_{UV}$ which has a part
linear in $\Lambda$ and another linear in $\ln \Lambda/\mu$.
\item \textbf{Introduction of couplings} $\lambda$, $g$ --- We
    expand the Wilson action in powers of fields to obtain
\begin{eqnarray}
\SI &=& \int_p \frac{1}{2} u_2 (\Lambda; p,-p)\, \phi (p) \phi (-p)\nn\\
&& + \int_{p_1,p_2,p_3} \frac{1}{4!} u_4 (\Lambda; p_1, p_2, p_3,
-p_1-p_2-p_3)\, \phi (p_1) \phi (p_2) \phi (p_3) \phi (p_4)\nn\\
&& + \int_p z(\Lambda; p^2) \frac{1}{2} \tilde{\chi}^I (-p) i \vec{p}
\cdot \vec{\sigma} \chi^I (p) \nn\\
&& + \int_{p,k} \frac{1}{\sqrt{N}} G(\Lambda; p,k) \phi (k)
\frac{1}{2} \tilde{\chi}^I (-p-k) \chi^I (p) + \cdots\,.
\end{eqnarray}
Choose a finite renormalization scale $\mu$.  We then impose
\begin{equation}
\label{parameters}
\lb\begin{array}{r@{~=~}l}
 u_2 (\mu; 0, 0)\Big|_{m^2=0} & 0\,,\\
 \frac{\partial}{\partial m^2} u_2 (\mu; 0,0)\Big|_{m^2=0} & 0\,,\\
 \frac{\partial}{\partial p^2} u_2 (\mu; p,-p)\Big|_{p^2=m^2=0} & 0\,,\\
 z(\mu; 0)\Big|_{m^2=0} & 0\,,\\
 u_4 (\mu; 0,0,0,0)\Big|_{m^2=0} & - \lambda\,,\\
 G (\mu; 0,0)\Big|_{m^2=0} & - g \,.
\end{array}\right.
\end{equation}
The first four are normalization conditions, and the last two
introduce coupling constants $\lambda$ \& $g$.  Note that the squared mass
parameter $m^2$ is introduced through $\SF$.
\end{enumerate}

The conditions (\ref{ERGdiff}-\ref{parameters}) determine $\SL$
uniquely as a functional of $\phi$ \& $\chi$; $\SL$ depends on the
parameters $m^2, \lambda, g$ and the mass scales $\Lambda, \mu$.
$\SL$ also depends on a particular choice of the cutoff function $K$,
but it can be shown formally that the dependence can be absorbed by
the normalization of parameters and fields. (For example, see Appendix
B.2 of Ref.~\citen{Igarashi:2009tj}.)

The beta functions and anomalous dimensions are obtained as the $\mu$
dependence of the Wilson action \cite{Pernici:1998tp,Sonoda:2006ai}:
\begin{equation}
- \mu \frac{\partial}{\partial \mu} \SL = \beta_m \mathcal{O}_m +
\beta_\lambda \mathcal{O}_\lambda + \beta_g \mathcal{O}_g +
\gamma_\phi \mathcal{N}_\phi + \gamma_\chi \mathcal{N}_\chi\,.
\label{mudep}
\end{equation}
The operators $\Op_m, \Op_\lambda, \Op_g$ are defined by
\begin{eqnarray}
\Op_m &\equiv& - \frac{\partial}{\partial m^2} \SL \nn\\
&&\, - \int_p \frac{\K{p} \left(1 - \K{p}\right)}{(p^2 + m^2)^2} 
\frac{1}{2} 
\lb \frac{\delta \SL}{\delta \phi (p)} \frac{\delta \SL}{\delta \phi
  (-p)} + \frac{\delta^2 \SL}{\delta \phi (p) \delta \phi (-p)} \rb\,,\\
\Op_\lambda &\equiv& - \frac{\partial}{\partial \lambda} \SL\,,\\
\Op_g &\equiv& - \frac{\partial}{\partial g} \SL\,.
\end{eqnarray}
These generate infinitesimal changes of the parameters $m^2, \lambda,
g$, respectively.  Denoting the $\Lambda$ independent correlation
functions by brackets, we obtain
\begin{equation}
\lb\begin{array}{c@{~=~}l}
\vev{\Op_m \,\phi (p_1) \cdots \chi^{I_1} (q_1) \cdots}_{m^2, \lambda,
  g} & - \partial_{m^2} \vev{\phi (p_1) \cdots \chi^{I_1} (q_1)
  \cdots}_{m^2, \lambda, g}\,,\\ 
\vev{\Op_\lambda \,\phi (p_1) \cdots\chi^{I_1} (q_1) \cdots}_{m^2, \lambda, g}
& - \partial_\lambda \vev{\phi (p_1) \cdots \chi^{I_1} (q_1)
  \cdots}_{m^2, \lambda, g}  \,, \\ 
\vev{\Op_g\, \phi (p_1) \cdots\chi^{I_1} (q_1) \cdots}_{m^2, \lambda,
  g} & - \partial_g \vev{\phi (p_1) \cdots \chi^{I_1} (q_1)
  \cdots}_{m^2, \lambda, g}  \,,
\end{array}\right.
\end{equation}
where the dots stand for a string of elementary fields $\phi$ and
$\chi^I$.  The operators $\mathcal{N}_\phi$ and $\mathcal{N}_\chi$ are
the equation-of-motion operators defined by
\begin{eqnarray}
\mathcal{N}_\phi &\equiv& - \int_p \K{p} \left(
[\phi] (p) \frac{\delta\SL}{\delta \phi (p)} + \frac{\delta}{\delta
  \phi (p)} [\phi] (p) \right)\,,\\
\mathcal{N}_\chi &\equiv& - \int_p \K{p} \left(
\SL \Rd{\chi^I (p)} [\chi^I] (p) - \Tr [\chi^I] (p) \Rd{\chi^I (p)} \right)\,,
\end{eqnarray}
where
\begin{eqnarray}
    \left[\phi\right] (p) &\equiv& \phi (p) + \frac{1 - \K{p}}{p^2 +
      m^2} \frac{\delta  \SI}{\delta \phi (-p)}\,,\\
    \left[\chi^I\right] (p) &\equiv& \chi^I (p) + \frac{1-\K{p}}{p^2}
    (-i) \vec{p} \cdot \vec{\sigma} \Ld{\tilde{\chi}^I (-p)} \SI\,.
\end{eqnarray}
They count the number of fields in the correlation functions:
\begin{equation}
\lb\begin{array}{c@{~=~}l}
\vev{\mathcal{N}_\phi\, \phi (p_1) \cdots \phi (p_n) \chi^{I_1} (q_1)
  \cdots \chi^{I_{2k}} (q_{2k})}_{m^2, \lambda, g} & n \vev{\phi (p_1)
  \cdots \chi^{I_1} (q_1) \cdots}_{m^2, \lambda, g} \,,\\ 
\vev{\mathcal{N}_\chi\, \phi (p_1) \cdots \phi (p_n) \chi^{I_1} (q_1)
  \cdots \chi^{I_{2k}} (q_{2k})}_{m^2, \lambda, g} & 2k \vev{ \phi (p_1)
  \cdots \chi^{I_1} (q_1) \cdots}_{m^2, \lambda, g} \,.
\end{array}\right.
\end{equation}

The meaning of (\ref{mudep}) is clear.  Its correlation with a product
of elementary fields gives the RG equation:
\begin{eqnarray}
&&\left( - \mu \frac{\partial}{\partial \mu} + 
\beta_m \frac{\partial}{\partial m^2} + \beta_g
    \frac{\partial}{\partial g} + \beta_\lambda
    \frac{\partial}{\partial \lambda} \right) \vev{\phi (p_1) \cdots
  \chi^{I_1} (q_1) \cdots}_{m^2, \lambda, g}\nn\\
&& \qquad = \left( n_\phi \gamma_\phi + n_\chi \gamma_\chi \right)
 \vev{\phi (p_1) \cdots
  \chi^{I_1} (q_1) \cdots}_{m^2, \lambda, g}\,,
\end{eqnarray}
where $n_\phi, n_\chi$ are the number of $\phi$'s and $\chi$'s in the
correlator.

In order to compute the beta functions at 1-loop, we need to compute
$\SI$ at 1-loop, in particular the coefficients $u_2, u_4, z, G$.  The
results are given in Appendix \ref{app-1loop}.  Taking the $\mu$
derivatives, we obtain the following results:
\begin{eqnarray}
\frac{1}{\mu^2} \beta_m &=&\frac{\lambda}{\mu}  \frac{1}{2} I_1 - 
\frac{g^2}{\mu} 2 I_2 - \frac{m^2}{\mu^2} \left( \frac{\lambda}{\mu}
    \frac{1}{2} I_4 + \frac{g^2}{\mu} \frac{1}{6} (I_3 + 2 I_5) \right)\,,\\ 
\frac{1}{\sqrt{\mu}} \beta_g &=& - \frac{g^3}{\mu^{\frac{3}{2}}} \lb
\frac{1}{2 N} (I_5 + 2 I_6) + \frac{1}{12} (I_3 + 2 I_5) \rb\,,\\
\frac{1}{\mu} \beta_\lambda &=& - \frac{\lambda^2}{\mu^2} \frac{3}{2}
I_5 + \frac{g^4}{\mu^2} \frac{6}{N} I_7 - \frac{\lambda g^2}{\mu^2}
\frac{1}{3} (I_3 + 2 I_5)\,,\\
\gamma_\phi &=& \frac{g^2}{\mu} \frac{1}{12} (I_3 + 2 I_5)\,,\\
\gamma_\chi &=& \frac{g^2}{\mu} \frac{1}{4 N} I_5\,,
\end{eqnarray}
where the integrals $I$'s are defined in terms of the cutoff function
$K$ and its derivative $\Delta$ in Appendix \ref{app-integrals}.

\section{RG equations and fixed points\label{RG equations}}

We have obtained 1-loop beta functions and anomalous dimensions.  To
obtain RG flows that describe the phase structure of the Yukawa model,
we need to rescale both space and fields so that the renormalization
scale $\mu$ is fixed under the RG flows.  Due to this resclaing, the
beta functions acquire contribution from the engineering dimensions of
the parameters.  Calling $\frac{m^2}{\mu^2}$ as $m^2$,
$\frac{\lambda}{\mu}$ as $\lambda$, and $\frac{g}{\sqrt{\mu}}$ as $g$,
the flow equations for these dimensionless parameters become
\begin{equation}
\lb\begin{array}{c@{~=~}l}
 \frac{d m^2}{dt} & 2 m^2 + \beta_m\,,\\
 \frac{d g^2}{dt} &  g^2 + 2 g \beta_g\,,\\
 \frac{d \lambda}{dt} & \lambda + \beta_\lambda\,.
\end{array}\right.
\end{equation}
These equations are valid in a neighborhood of the origin $m^2 = g^2 =
\lambda = 0$, called the Gaussian fixed point.  As has been explained in
\S\ref{introduction}, we restrict ourselves only to the region of
parameters accessible from the Gaussian fixed point.  In other words, we
only follow the flows that originate from the origin.  All the
non-trivial fixed points we will discuss in this section are accessible
from the Gaussian fixed point.

At 1-loop, using the results of the previous section, we obtain
\begin{equation}
\lb\begin{array}{c@{~=~}l}
\frac{d}{dt} m^2  & \lb 2 - \lambda \frac{1}{2} I_4 -
g^2 \frac{1}{6} \left( I_3 + 2 I_5 \right) \rb
m^2 + \lambda  \frac{1}{2} I_1 - g^2 2 I_2\,,\\
\frac{d}{dt} g^2 &  g^2 \left( 1 - \frac{g^2}{g_*^2} \right)\,,\\
\frac{d}{dt} \lambda & \frac{1}{\lambda_I} \left(
\lambda_+ (g^2) - \lambda \right) \left(
\lambda - \lambda_- (g^2) \right)\,,
\end{array}\right.
\end{equation}
where the integrals $I$'s are defined in terms of the cutoff function
$K$ in Appendix \ref{app-integrals}, $g_*^2$ and $\lambda_I$ by
\begin{equation}
\lb\begin{array}{c@{~\equiv~}l}
\frac{1}{2 g_*^2} & \frac{1}{12} \left( I_3 + 2 I_5\right) + \frac{1}{2 N}
\left( I_5 + 2 I_6\right)\,,\\
\frac{1}{\lambda_I} & \frac{3}{2} I_5\,,
\end{array}\right.
\end{equation}
and $\lambda_{\pm} (g^2)$ by
\begin{equation}
\lb\begin{array}{c@{~=~}l}
\lambda_+ (g^2) + \lambda_- (g^2) & \lambda_I \lb 1 - g^2 \frac{1}{3}
(I_3 + 2 I_5) \rb\,,\\
\lambda_+ (g^2) \lambda_- (g^2) & - \lambda_I g^4 \frac{6}{N} I_7\,.
\end{array}\right.
\end{equation}
With the convention $\lambda_+ (g^2) > \lambda_- (g^2)$, we obtain
\begin{equation}
\lambda_\pm (g^2) = 
 \frac{\lambda_I}{2} \left[ 1 - g^2 \frac{1}{3} (I_3
    + 2 I_5) \pm \sqrt{
  \left( 1 - g^2 \frac{1}{3} (I_3 + 2 I_5)\right)^2 + 
  \frac{g^4}{\lambda_I}  \frac{24}{N} I_7 }\,\, \right]\,. \label{lambda-star}
\end{equation}
Similarly, the anomalous dimensions are given by
\begin{equation}
\lb\begin{array}{c@{~=~}l}
 \gamma_\phi (g^2) & g^2 \frac{1}{12} \left( I_3 + 2 I_5 \right)\,,\\
 \gamma_\chi (g^2) & g^2 \frac{1}{4N} I_5\,,
\end{array}\right.
\end{equation}
which are independent of $\lambda$.

\begin{figure}[t]
\begin{center}
\includegraphics[scale=1.2]{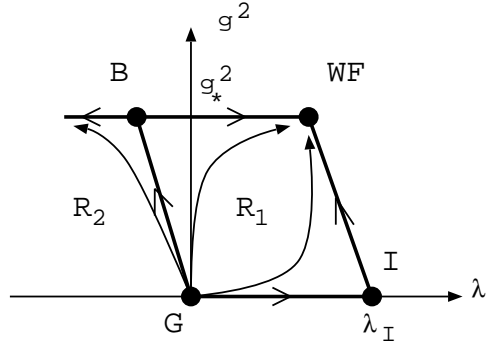}
\end{center}
\caption{\label{RGflows}Schematic RG flows. G for
  the Gaussian, I for the Ising, WF for the Wilson-Fisher, and B for
  the bicritical fixed points. The flow of $m^2$ is suppressed.  We
  have a continuous phase transition in Region 1, and a first order
  transition in Region 2.}
\end{figure}

The RG flows are given schematically in Fig.\,\ref{RGflows}.
For any $N = 1,2,\cdots$, the flows have four fixed points.
\begin{enumerate}
\item \textbf{Gaussian} --- This exists by construction; in this paper
    we only study the RG flows out of this fixed point.  At the
    Gaussian fixed point, all parameters vanish:
\begin{equation}
m^2 = \lambda = g = 0\,.
\end{equation}
\item \textbf{Ising} --- With $g=0$, the fermions decouple, and we get
    a $\phi^4$ theory.  Its fixed point corresponds to the critical
    Ising model.
\begin{equation}
\lb\begin{array}{c@{~=~}l}
m^2 & m_I^2 \equiv - \frac{\lambda_I \frac{I_1}{2}}{2 - \lambda_I
  \frac{I_4}{2}} = - \frac{I_1}{6 I_5 - I_4}\,,\\
\lambda & \lambda_I\,,\\
g^2 & 0\,.
\end{array}\right.
\end{equation}
The small deviation $\Delta m^2 \equiv m^2 - m_I^2$ has the scale
dimension
\begin{equation}
y_E \equiv 2 + \frac{\partial}{\partial m^2} \beta_m = 2 - \lambda_I
\frac{I_4}{2} = 2 - \frac{1}{3} \frac{I_4}{I_5}\,.
\end{equation}
Similarly, $\Delta \lambda \equiv \lambda - \lambda_I$ has the scale
dimension $-1$.  $\Delta m^2$ and $g^2$ are the two relevant
parameters at this fixed point.
\item \textbf{Wilson-Fisher} --- This is the most stable fixed point
    given by
\begin{equation}
\lb\begin{array}{c@{~=~}l}
 m^2 & m_{*+}^2 \equiv - \frac{1}{y_{E+}} \left( \lambda_{*+}
     \frac{I_1}{2} - g_*^2 2 I_2 \right) \,,\\
 \lambda & \lambda_{*+} \equiv \lambda_+ (g_*^2)\,,\\
 g^2 & g_*^2\,,
\end{array}\right.
\end{equation}
where $\lambda_{*+}$ is given explicitly by (\ref{lambda-star}).
The scale dimension of $\Delta m^2$ is given by
\begin{equation}
y_{E+} \equiv 2 - \lambda_{*+} \frac{I_4}{2} - g_*^2 \frac{1}{6} \left(
    I_3 +  2 I_5 \right)\,.
\end{equation}
$\Delta m^2$ is the only relevant parameter.  $\Delta \lambda$ is
irrelevant with the scale dimension
\begin{equation}
y_{\lambda +} \equiv - \frac{1}{\lambda_I} \left( \lambda_{*+} -
    \lambda_{*-} \right) < 0\,.
\end{equation}
\item \textbf{Bicritical} --- This has two relevant parameters, and we
    call this a bicritical fixed point.
\begin{equation}
\lb\begin{array}{c@{~=~}l}
  m^2 & m_{*-}^2 \equiv - \frac{1}{y_{E-}} \left( \lambda_{*+}
      \frac{I_1}{2} - g_*^2 2 I_2 \right)\,,\\
 \lambda & \lambda_{*-} \equiv \lambda_- (g_*^2)\,,\\
 g^2 & g_*^2\,,
\end{array}\right.
\end{equation}
where $y_{E-}$, the scale dimension of $\Delta m^2$, is given by
\begin{equation}
y_{E-} \equiv 2 - \lambda_{*-} \frac{I_4}{2} - g_*^2 \frac{1}{6} \left( I_3
    + 2 I_5 \right) \,.
\end{equation}
Unlike the Wilson-Fisher fixed point, $\Delta \lambda$ is relevant
with the scale dimension
\begin{equation}
y_{\lambda -} \equiv - y_{\lambda +} > 0\,.
\end{equation}
\end{enumerate}

\begin{table}[t]
\begin{center}
\caption{\label{exponents} Notation for critical exponents}
\begin{tabular}{l|c|c|c|c}
\hline
parameter&Gauss& Ising& Wilson-Fisher& Bicritical\\
\hline
$\Delta m^2$& $2$& $y_I>0$& $y_{E+}>0$& $y_{E-}>0$\\
$\Delta \lambda$& $1$& $-1$& $y_{\lambda +}<0$& $y_{\lambda
  -}=-y_{\lambda +} >0$\\
$\Delta g$& $\frac{1}{2}$& $\frac{1}{2}$& $-1$& $-1$\\
\hline
\end{tabular}
\end{center}
\end{table}

The critical exponents are universal, and accordingly they do not
depend on what cutoff function $K$ we use to formulate the ERG
differential equations.\cite{Latorre:2000qc} (See also Appendix B.2 of
Ref.~\citen{Igarashi:2009tj}.) Approximate solutions of the ERG
differential equations, whether they are perturbative or
non-perturbative, lose universality, and the critical exponents become
dependent on the choice of a cutoff function.  The cutoff dependence
is an inevitable artifact of the introduction of an approximation.

To compute the critical exponents numerically, we have used a
particular class of $K$ (See Fig.~\ref{app-Ka} in Appendix
\ref{app-integrals}):
\begin{equation}
K(x) = \lb \begin{array}{c@{~\quad~}l}
 1 & (x^2 < a^2)\\
 \frac{1 - x^2}{1 - a^2}& (a^2 < x^2 < 1)\\
 0& (x^2 > 1)\,.
\end{array}\right.
\end{equation}
The advantage of this cutoff function is that the necessary 1-loop
integrals, listed in Appendix \ref{app-integrals}, can be performed
analytically.  The cutoff function interpolates two popular choices,
one at $a = 0$ (Litim's cutoff \cite{Litim:2000ci}) and another at
$a=1$ (sharp cutoff \cite{Wegner:1972ih}).  The dependence of the
critical exponents on the parameter $a$ indicates the accurracy of the
numerical values obtained.

\begin{figure}
\begin{center}
\includegraphics[scale=0.5]{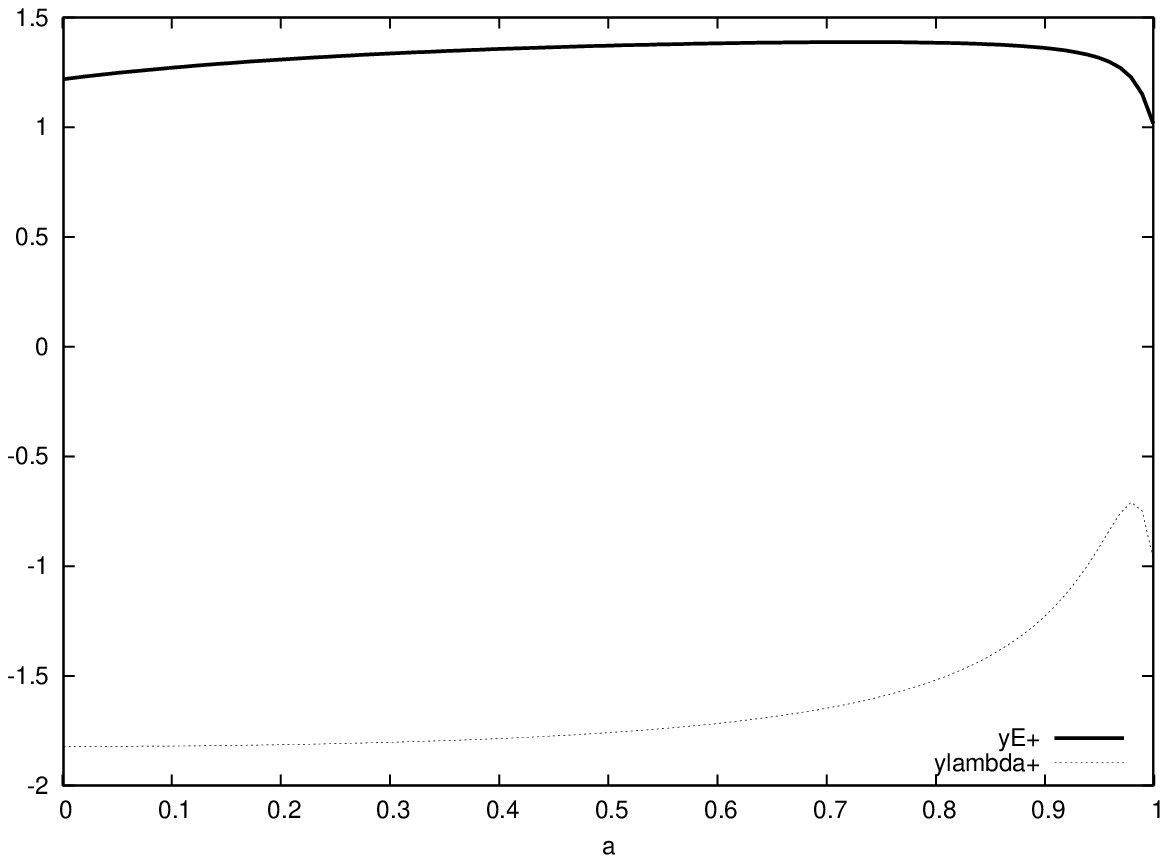}
\includegraphics[scale=0.5]{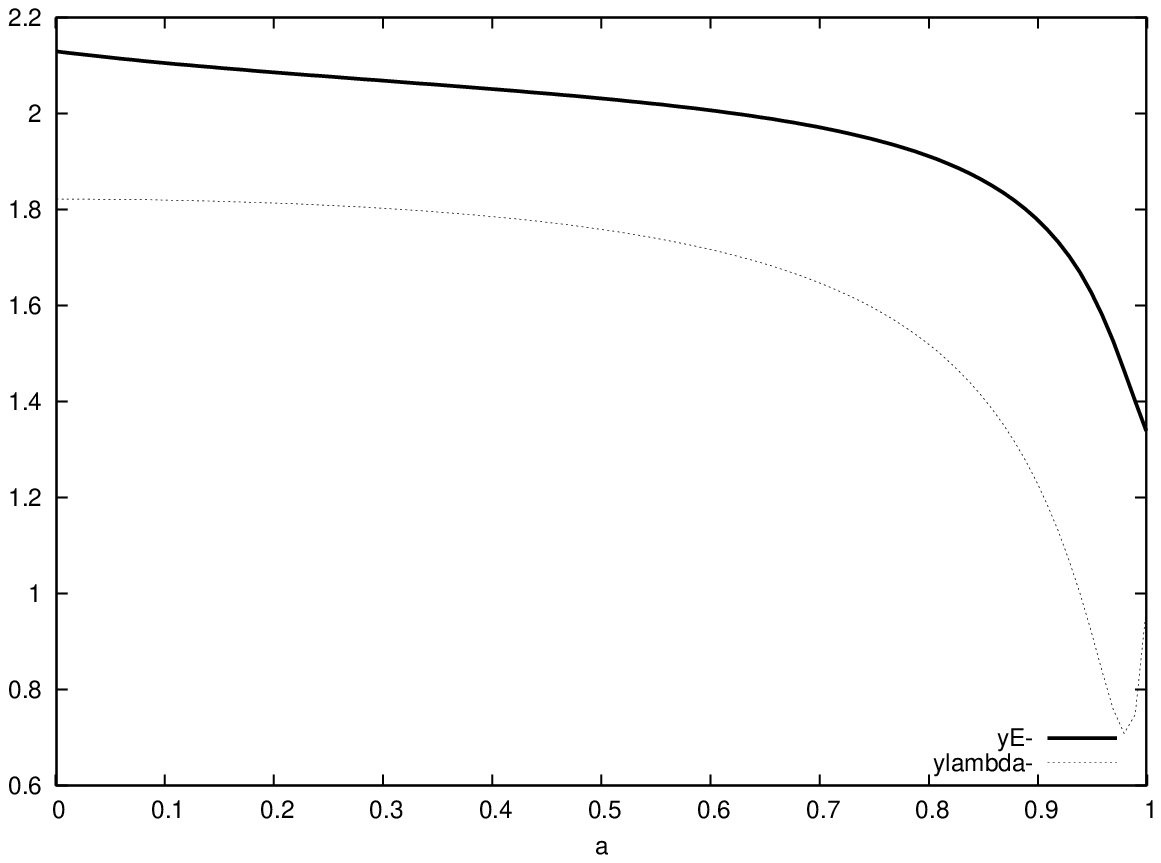}
\caption{\label{plot-exponents} Dependence of $y_{E \pm}, y_{\lambda
    \pm}$ on $a$ for $N=1$.  We note $y_{\lambda +} = - y_{\lambda
    -}$.}
\end{center}
\end{figure}

\begin{figure}[b]
\begin{center}
\includegraphics[scale=0.6]{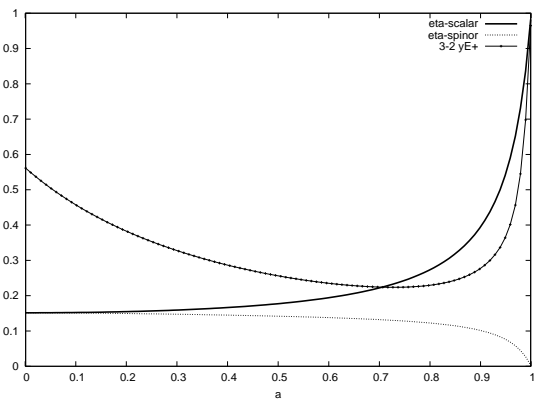}
\caption{\label{plot-etas} Dependence of $\eta_\phi, \eta_\chi$ on $a$
for $N=1$.  Supersymmetry would imply $\eta_\phi = \eta_\chi = 3- 2
y_{E+}$.  (See \S \ref{Comparison}.)}
\end{center}
\end{figure}

Let us plot the $a$-dependence of the critical exponents for $N=1$.
(Fig.~\ref{plot-exponents}) We find that the exponents do not have
much $a$-dependence for $0 < a < 0.8$.  $y_{E+}$ takes a maximum value
$1.39$ at $a \simeq 0.73$, $y_{\lambda+}$ a maximum $-0.71$ at $a =
0.98$, and $y_{E -}$ a maximum $1.82$ at $a=0$.  Since $y_{\lambda-} =
-y_{\lambda+}$, $y_{\lambda-}$ takes a minimum $0.71$ at $a = 0.98$.
We take these extremum values as our numerical estimates for the
exponents. \cite{Stevenson:1981vj} Note that the following relations
are satisfied:
\begin{equation}
y_{E+} > y_{\lambda +},\quad y_{E-} > y_{\lambda -}\,.
\end{equation}
Hence, at the bicritical fixed point, $\Delta m^2$ is more relevant
than $\Delta \lambda$.  We next plot the
$a$-dependence of the anomalous dimensions (Fig.~\ref{plot-etas}):
\begin{equation}
\eta_\phi \equiv 2 \gamma_\phi (g_*^2),\quad
\eta_\chi \equiv 2 \gamma_\chi (g_*^2)\,.
\end{equation}
At 1-loop, the Wilson-Fisher and bicritical fixed points share the same
anomalous dimensions. Again, the dependence on $a$ is mild in the
region $0 < a < 0.8$.  The extremum values at $a=0$ are
\begin{equation}
\eta_\phi = \eta_\chi = \frac{5}{33} \simeq 0.15\,.
\end{equation}

\section{Phase transitions\label{Phase transitions}}

Given $\lambda$ \& $g^2$, we examine the dependence of the model
on the squared mass parameter $m^2$ in this section.

Suppose $(\lambda, g^2)$ is in Region 1, surrounded by four RG flows
G-I, I-WF, B-WF, and G-B in Fig.~\ref{RGflows}.  Then the
two-dimensional flow is always attracted to the Wilson-Fisher fixed
point.  Let $m^2_{cr} (\lambda, g^2)$ be the value of $m^2$ such that
the RG flow starting from $(\lambda, g^2, m^2 = m_{cr}^2 (\lambda,
g^2))$ flows to the Wilson-Fisher fixed point as $t \to \infty$:
\begin{equation}
\lb
\begin{array}{c@{~\stackrel{t \to \infty}{\longrightarrow}~}l}
\lambda & \lambda_{*+}\,,\\
g^2 & g_*^2\,,\\
m^2_{cr}& m_{*+}^2\,.
\end{array}\right.
\end{equation}
Since $y_{E+} > 0$, the deviation $\Delta m^2 \equiv m^2 - m_{*+}^2$
is relevant, and grows along the RG flow.  Hence, the model exhibits a
continuous phase transition at $m^2 = m_{cr}^2 (\lambda, g^2)$.  We
expect that the $\mathbf{Z_2}$ is exact for $m^2 > m_{cr}^2$ and
broken for $m^2 < m_{cr}^2$.

Let us now suppose $(\lambda, g^2)$ is in Region 2, to the left of the
RG flow G-B in Fig.~\ref{RGflows}.  The two-dimensional flow has no
fixed point to reach, and this implies the existence of a transition
point $m^2_{tr} (\lambda, g^2)$ that exhibits a first order phase
transition.  As for Region 1, we expect that the $\mathbf{Z_2}$ is
exact for $m^2 > m_{tr}^2$ and broken for $m^2 < m_{tr}^2$.
In Appendix \ref{app-transitions} we compute $m_{cr}^2$ and $m_{tr}^2$
by solving the 1-loop RG equations analytically.  (For the first order
transition, only the infinitesimal region below G-B is considered.)

\begin{figure}
\begin{center}
\includegraphics{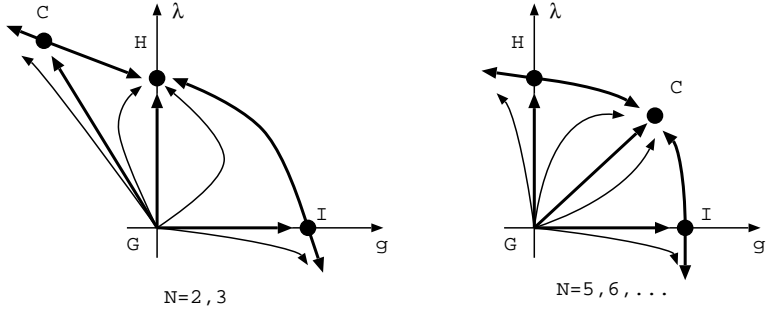}
\end{center}
\caption{\label{N-vector}RG flows of the N-vector model with cubic
  anisotropy in $3$ dimensions --- G, H, I stand for Gauss,
  Heisenberg, and Ising, respectively.  First order transitions take
  place in the regions below G-I and on the left of G-C ($N\le 4$) or
  G-H ($N\ge 4$). H and C merge at $N=4$.}
\end{figure}

The phase structure of the Yukawa model discussed above is similar to
that of the N-vector model with cubic anisotropy in $3$
dimensions. \cite{PhysRevB.18.1406} (See also \S 5.8.5 of
Ref.~\citen{Chaikin:1995}, for example.)  The classical lagrangian of the
model is given by
\begin{equation}
\mathcal{L} = \frac{1}{2} \sum_{I=1}^N \lb (\nabla \phi_I)^2 + m^2
\phi_I^2\rb + \frac{\lambda}{8} \big( \sum_{I=1}^N \phi_I^2\big)^2 +
\frac{g}{4!} \sum_{I=1}^N \phi_I^4\,,
\end{equation}
where $N \ge 2$.  The phase structure depends on $N$, but it is
similar to the phase structure of the Yukawa model found here with the
presence of a region of first order transitions.
(Fig.~\ref{N-vector}) 

\section{Comparison of the $N=1$ case with the Wess-Zumino model\label{Comparison}} 

The $N=1$ case is particularly interesting since the model belongs to
the same universality class as the $\mathcal{N} = 1$ supersymmetric
Wess-Zumino model.  The Wess-Zumino model has recently been studied
with ERG, \cite{Synatschke:2010ub} and we would like to make sure that
our results are compatible.

The Wilson-Fisher fixed point is characterized by three critical
exponents: $y_{E+}$, $\eta_\phi$, and $\eta_\chi$.  Supersymmetry
implies
\begin{equation}
\eta_\phi = \eta_\chi\,.
\end{equation}
As Fig.~\ref{plot-etas} shows, this relation is satisfied at the
extremum $a=0$, and reasonably satisfied for $a \in [0, 0.8]$.  An
interesting relation
\begin{equation}
3 - 2 y_{E+} = \eta_\phi = \eta_\chi
\end{equation}
found in Ref.~\citen{Synatschke:2010ub} is also reasonably satisfied
for $a \in [0.6,0.8]$.  (Fig.~\ref{plot-etas}) Considering the
crudeness of our 1-loop approximations, the agreement is satisfactory.
\begin{table}[t]
\begin{center}
\caption{\label{N=1 exponents}Numerical estimates for the critical
  exponents at the $N=1$ Wilson-Fisher fixed point}
\begin{tabular}{c|l|l|l|l|l}
\hline
& $y_{E+}$& $y_{\lambda +}$& $\eta_\phi$& $\eta_\chi$& $3 - 2 y_{E+}$\\
\hline
Yukawa& $1.39$& $-0.71$& $0.15$& $0.15$& $0.22$\\
Wess-Zumino \cite{Synatschke:2010ub}& $1.406$ & $-0.756$ &$0.188$ &
$0.188$ & $0.188$\\ 
\hline
\end{tabular}
\end{center}
\end{table}
In Appendix \ref{app-WZ} we apply our perturbative ERG formalism
directly to the WZ model and compute the critical exponents.  The
agreement with Ref.~\citen{Synatschke:2010ub} improves slightly.

\section{Comparison with the Gross-Neveu model\label{GN}}

\begin{figure}
\includegraphics[scale=0.5]{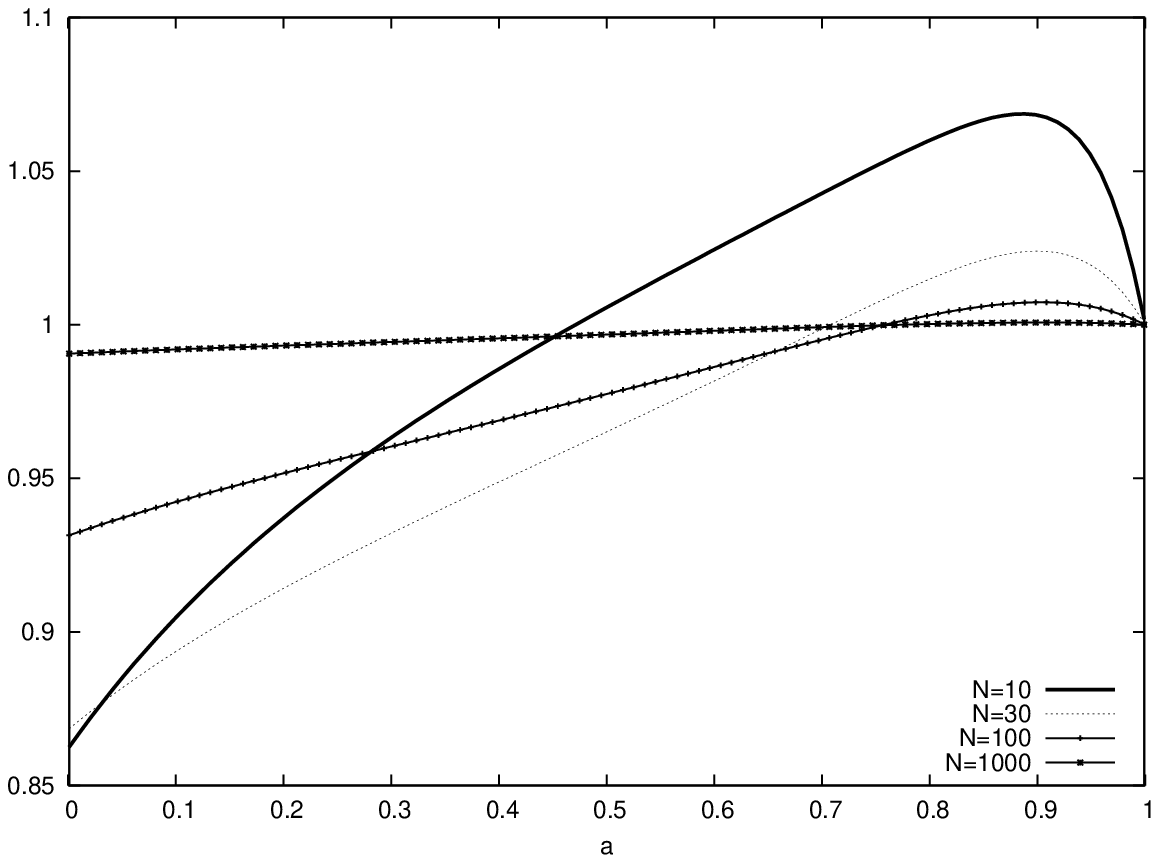}
\includegraphics[scale=0.5]{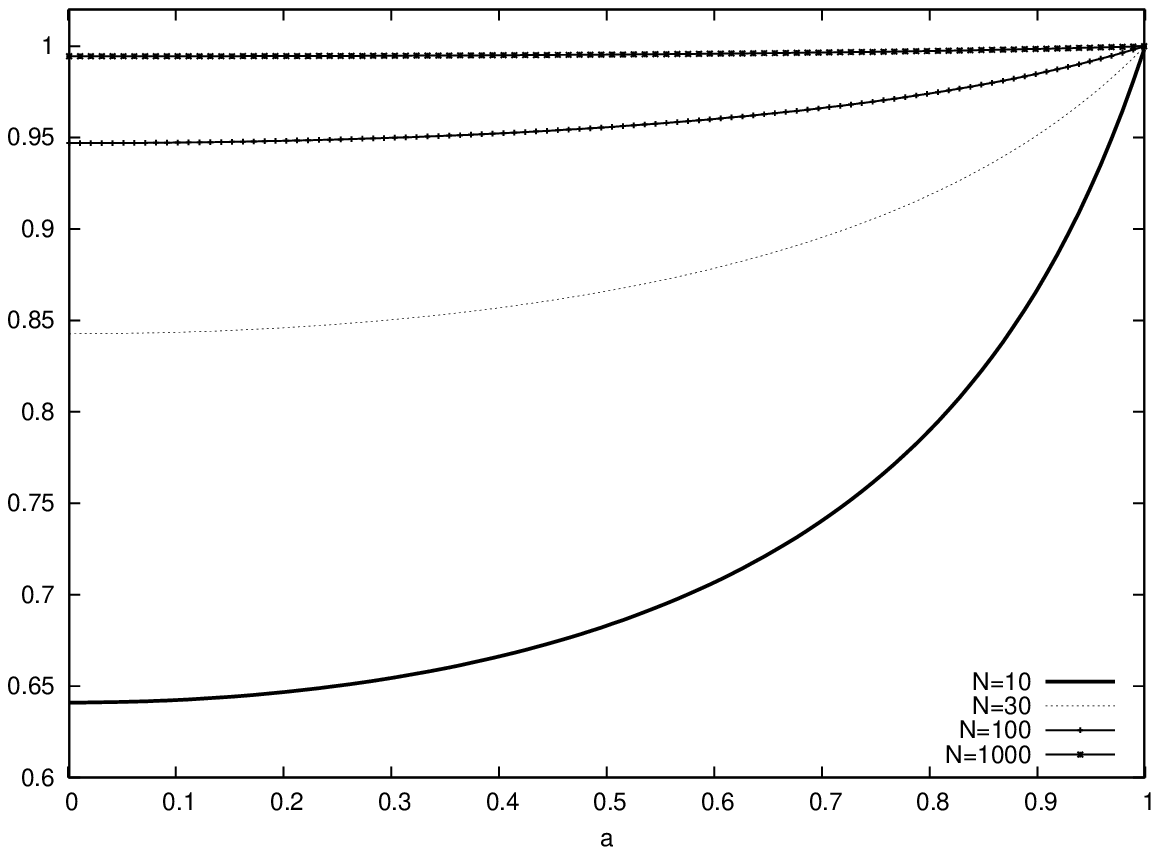}
\caption{\label{plot-Ndep}$N$ dependence of $y_{E+}$ (left) and $\eta_\phi$
  (right).  The $N$ dependence suggests that the preferred values of
  $a$ are given by the peak near $a=1$ for $y_{E+}$, and $a=0$ for $\eta_\phi$.}
\end{figure}

Let us consider the large $N$ limit of the RG equations of sect.~\ref{RG
  equations}, which gives
\begin{equation}
\lb\begin{array}{c@{~=~}l@{\quad}c@{~=~}l}
 y_{E+} & 1,&
 y_{E-} & \frac{4}{3}\,,\\
 \eta_{\phi}& 1,&
 \eta_{\chi}& 0\,.
\end{array}\right.
\end{equation}
The Gross-Neveu model with $\frac{N}{2}$ complex fermions has been
studied with ERG in \citen{Rosa:2000ju}, where the RG flow of the
Yukawa model was run numerically with the initial condition
corresponding to the Gross-Neveu model.  In TABLE \ref{comparison
  large N} the results of Ref.~\citen{Rosa:2000ju} are compared with
ours for $y_{E+}$, evaluated at a peak near $a=1$, and $\eta_\phi$,
evaluated at $a=0$.  (Fig.~\ref{plot-Ndep})
\begin{table}[t]
\caption{\label{comparison large N}Numerical estimates of critical
  exponents for large $N$.  Our results are compared with those
from Ref.~\citen{Rosa:2000ju}.  (In \citen{Rosa:2000ju}, 
the notations $\nu = 1/y_{E+}$, $\eta_\sigma = \eta_\phi$ are used.)}
\begin{center}
\begin{tabular}{c|l|l|l|l|l|l|l|l}
\hline
&\multicolumn{2}{c|}{$N=4$}& \multicolumn{2}{c|}{$N=6$}&
\multicolumn{2}{c|}{$N=8$}& \multicolumn{2}{c}{$N=24$}\\ 
\hline
&$y_{E+}$& $\eta_\phi$& $y_{E+}$& $\eta_\phi$& $y_{E+}$& $\eta_\phi$&
$y_{E+}$& $\eta_\phi$\\
\hline
ours&$1.16$& $0.42$ & $1.11$& $0.52$&$1.08$& $0.59$&$1.03$&$0.81$\\
Ref.~\citen{Rosa:2000ju}&$1.041$& $0.561$&$0.961$& $0.710$&$0.990$&
$0.789$&$0.978$&$0.936$ \\
\hline
\end{tabular}
\end{center}
\end{table}

In Ref.~\citen{Hands:1991py}, the critical exponents have been calculated
to order $\frac{1}{N}$ as follows:
\begin{equation}
\lb
\begin{array}{c@{~=~}l}
 y_{E+} & 1 - \frac{16}{3 \pi^2} \frac{1}{N}\,,\\
 \eta_\phi & 1 - \frac{32}{3 \pi^2} \frac{1}{N}\,.
\end{array}\right.
\end{equation}
(We have replaced $N$, the number of complex fermions, in
Ref.~\citen{Hands:1991py} by $\frac{N}{2}$, where $N$ is the number of
real fermions.) Our results reproduce the correct large N limit, but
not the $\frac{1}{N}$ corrections.  The most serious failure of our
1-loop approximation is the wrong sign for the $\frac{1}{N}$
correction to $y_{E+}$; in comparison the correct sign has been
obtained in \citen{Rosa:2000ju}.

\begin{figure}[b]
\begin{center}
\includegraphics{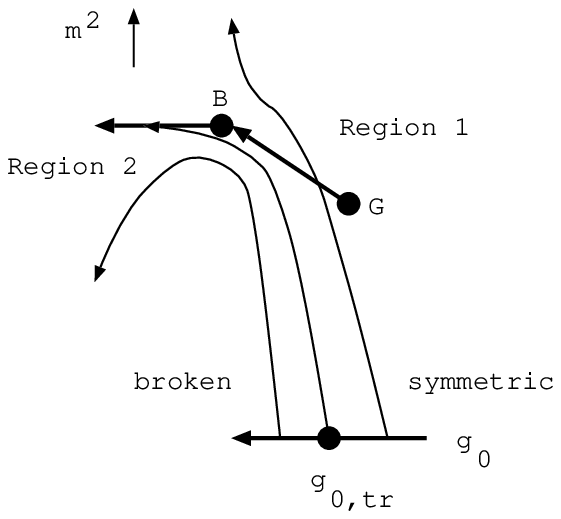} \includegraphics{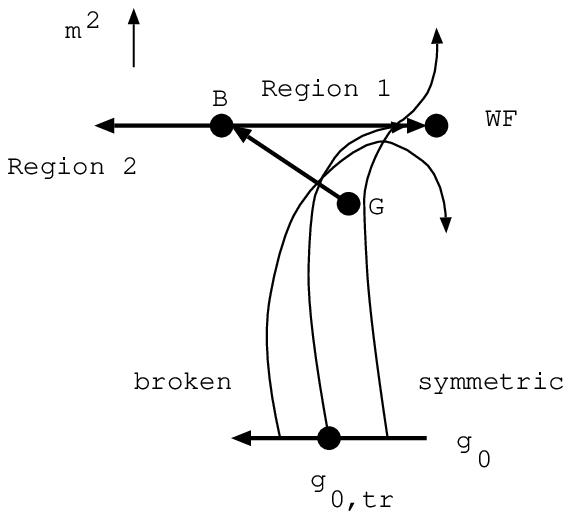}
\end{center}
\caption{\label{N=2 GN}The flow (left) starting from the transition point
  $g_{0, tr}$ of the $N=2$ Gross-Neveu model does not flow to a fixed
  point.  This is compared with the flow (right) if the transition were
  continuous.}
\end{figure}

We note that a first order transition was found for $N=2$ in
Ref.~\citen{Rosa:2000ju}.  This can be explained if we assume that the
transition point of this model belongs to Region 1 of the $(\lambda,
g^2)$ parameter space. (Fig.~\ref{N=2 GN})

\section{Concluding remarks\label{Conclusions}}

In this paper we have applied the ERG formalism perturbatively to the
Yukawa model in three dimensions to obtain the RG flows.  We have
found a phase structure similar to that of the N-vector model with
cubic anisotropy.  Accordingly, the spontaneous breaking of the
$\mathbf{Z_2}$ symmetry of the Yukawa model can take place either at a
continuous or at a first order transition point.  The existence of a
domain of parameters for the first order transition should be verified
further, perhaps, by studying the effective potential with ERG.

The Yukawa model with one real spinor includes the $\mathcal{N}=1$
Wess-Zumino model as a subset, and the fixed point of the latter is
inevitably that of the former.  At the Wilson-Fisher type fixed point of
the Yukawa model, the only relevant parameter preserves supersymmetry;
hence, the long distance behavior of an almost critical Yukawa model is
the same as a massive Wess-Zumino model with its supersymmetry either
exact or broken spontaneously.  Emergence of supersymmetry at a critical
point of a non-supersymmetric theory has also been found for the Yukawa
model with complex scalars and spinors. \cite{Lee:2010fy}

The way we apply ERG perturbatively, we can only study the RG flows of
UV renormalizable theories.  This constraint still leaves a wide class
of models as an object of study, and we advocate further perturbative
applications of ERG for its simplicity and good cost-performance
ratio.

\section*{Acknowledgements}

I thank the organizers and participants of the ERG2010 conference
    held in Corfu, Greece for a stimulating atmosphere where partial
    results of this paper were presented.  I also thank Yannick
    Meurice for critical comments and for Ref.~\citen{Lee:2010fy}.
    This work was partially supported by JSPS Grant-In-Aid \#22540282.

\appendix
%

\section{Spinors in $D=3$\label{app-spinors}}

For the reader's convenience, we summarize the salient features of
spinors in three dimensional Minkowski and Euclidean spaces.

\subsection{Minkowski space}

We can choose the gamma matrices pure imaginary:
\begin{equation}
\gamma^0 = \sigma_y,\quad \gamma^1 = i \sigma_x,\quad \gamma^2 = i
\sigma_z\,.
\end{equation}
Under a Lorentz transformation, a two-component spinor $\chi$
transforms as
\begin{equation}
\chi \longrightarrow A \chi\,,
\end{equation}
where $A \in SL(2,R)$ is a 2-by-2 real matrix with determinant $1$.
Since $A$ is real, we can assume $\chi$ to be real.  We define
\begin{equation}
\tilde{\chi} \equiv \chi^T \sigma_y
\end{equation}
which transforms as
\begin{equation}
\tilde{\chi} \longrightarrow \tilde{\chi} A^{-1}\,,
\end{equation}
since
\begin{equation}
\sigma_y A^T \sigma_y = A^{-1}\,.\quad (A \in SL(2,R))
\end{equation}

The lagrangian of the Yukawa model is given by
\begin{equation}
\mathcal{L} = \frac{1}{2} \partial^\mu \phi \partial_\mu \phi -
\frac{m^2}{2} \phi^2 + \frac{1}{2} \tilde{\chi}^I 
i \gamma^\mu \partial_\mu \chi^I - g \phi \frac{1}{2} \tilde{\chi}^I
\chi^I - \frac{\lambda}{4!} \phi^4\,.
\end{equation}
This is invariant under the parity transformation defined by
\begin{equation}
\lb\begin{array}{c@{~\longrightarrow~}l}
 \phi (x,y,t) & - \phi (-x,y,t)\,,\\
 \chi (x,y,t) & - i \gamma^1 \chi (-x,y,t) = \sigma_x \chi (-x,y,t)\,,\\
 \tilde{\chi} (x,y,t) & \tilde{\chi} (-x,y,t) i \gamma^1 = 
 \tilde{\chi} (-x,y,t) (-\sigma_x)\,.
\end{array}\right.
\end{equation}
The mass term $\tilde{\chi} \chi$ is forbidden by this invariance.

A Dirac spinor is a linear combination of two real spinors:
\begin{equation}
\Psi = \frac{1}{\sqrt{2}} \left(\psi + i \psi'\right)\,.
\end{equation}
We find
\begin{equation}
\bar{\Psi} \equiv \Psi^\dagger \gamma^0 = \frac{1}{\sqrt{2}}
\left(\tilde{\psi} - i \tilde{\psi}'\right)\,.
\end{equation}
We obtain
\begin{equation}
\bar{\Psi} \left( i \gamma^\mu \partial_\mu - M \right) \Psi =
\frac{1}{2} \left(\tilde{\psi} (i \gamma^\mu \partial_\mu - M) \psi +
\tilde{\psi}' (i \gamma^\mu \partial_\mu - M) \psi' \right)\,.
\end{equation}
A Dirac spinor can be obtained by the dimensional reduction of a Weyl
(or Majorana) spinor in $(3+1)$-dim Minkowski space.

\subsection{Euclidean space}

We can choose
\begin{equation}
\vec{\gamma} = \vec{\sigma}
\end{equation}
corresponding to the spin $\frac{1}{2}$ representation, familiar from
non-relativistic quantum mechanics.  Under a space rotation, a
two-component spinor transforms as
\begin{equation}
\psi \longrightarrow U \psi\,,
\end{equation}
where $U \in SU(2)$.  $\tilde{\psi} \equiv \psi^T \sigma_y$ transforms
as
\begin{equation}
\tilde{\psi} \longrightarrow \tilde{\psi} U^{-1}\,.
\end{equation}
Hence, the lagrangian
\begin{equation}
\mathcal{L} = \frac{1}{2} \left( \tilde{\psi} \vec{\sigma} \cdot
    \nabla \psi + m \tilde{\psi} \psi \right)
\end{equation}
is invariant under $SU(2)$.  

We call $\psi$ a real fermion, and call
a linear combination of two real fermions $\psi, \psi'$
\begin{equation}
\Psi = \frac{1}{\sqrt{2}} \left(\psi + i \psi'\right)
\end{equation}
a complex fermion.  Denoting $\bar{\Psi} = \frac{1}{\sqrt{2}}
(\tilde{\psi} - i \tilde{\psi}' )$, we can write the lagrangian for
$\Psi$ as
\begin{equation}
\mathcal{L} = \bar{\Psi} \left( \vec{\sigma} \cdot \nabla + m \right)
\Psi\,.
\end{equation}
$\Psi$ and $\bar{\Psi}$ are two independent 2-component Grassmann
fields.

\section{Wilson action at 1-loop\label{app-1loop}}

Defining
\begin{equation}
\Delta (q) \equiv - 2 q^2 \frac{d}{dq^2} K(q),\quad
\tilde{\Delta} (q) \equiv - 2 q^2 \frac{d}{dq^2} \Delta (q)\,,
\end{equation}
the 1-loop vertices are calculated as follows:
\begin{enumerate}
\item \underline{scalar 2-point}
\begin{eqnarray}
&&u_2 (\Lambda; p,-p) = \frac{\lambda}{2} \left[(\Lambda - \mu) \int_q
\frac{\Delta (q)}{q^2} + m^2 \lb \int_q \frac{1-\K{q}}{q^2 (q^2+m^2)}
- \frac{1}{\mu} \int_q \frac{1-K(q)}{q^4} \rb \right]\nn\\
&&\quad + g^2 \left[ \frac{p^2}{2} \left( - \int_q \frac{1 - \K{q}}{q^2}
        \frac{1 - \K{(q+p)}}{(q+p)^2} + \frac{1}{\mu}\int_q \left(
            \frac{1-K(q)}{q^2} \right)^2 \right)\right.\nn\\
&&\quad\qquad + \int_q \frac{1}{q^2} \lb \left(1 - \K{q}\right) \left( 1-
    \K{(q-p)}\right) - \left(1 - \K{q}\right)^2 \rb \nn\\
&&\quad\qquad + \frac{c_2}{\mu}  p^2
\left. - 2(\Lambda-\mu) \int_q \frac{1}{q^2} \Delta (q) (1-K(q))\right]\,,
\end{eqnarray}
where the constant $c_2$ is given by
\begin{equation}
c_2 = - \frac{1}{6} \int_q \frac{1}{q^4} (1-K(q)) (\Delta (q) -
\tilde{\Delta} (q))\,.
\end{equation}
\item \underline{fermion 2-point}
\begin{eqnarray}
&&z_2 (\Lambda; p^2)  
= \frac{g^2}{N} \left[ - \int_q \frac{1-\K{(q+p)}}{(q+p)^2} \frac{1
      - \K{q}}{q^2 + m^2} + \frac{1}{\mu} \int_q
    \left(\frac{1-K(q)}{q^2}\right)^2 \right.\nn\\
&&\left.- \int_q \lb \frac{1 - \K{(q+p)}}{(q+p)^2} \frac{1 - \K{q}}{q^2 +
  m^2} - \left( \frac{1 - \K{q}}{q^2} \right)^2 \rb \frac{q p}{p^2} +
\frac{k_2}{\mu} \right] \,,
\end{eqnarray}
where the constant $k_2$ is given by
\begin{equation}
k_2 = \frac{1}{3} \int_q \frac{1}{q^4} (1-K(q)) \lb - 2 (1-K(q)) +
  \Delta (q) \rb\,.
\end{equation}
\item \underline{Yukawa coupling}
\begin{equation}
G(\Lambda; 0,0) = - g + \frac{g^3}{N} \left[
\int_q \frac{\left(1 - \K{q}\right)^3}{q^2 (q^2 + m^2)} -
\frac{1}{\mu} \int_q \frac{(1-K(q))^3}{q^4} \right]\,.
\end{equation}
\item \underline{scalar 4-point}
\begin{eqnarray}
u_4 (\Lambda; 0,0,0,0) &=& - \lambda + \frac{3 \lambda^2}{2} \int_q \lb
\left( \frac{1 - \K{q}}{q^2 + m^2}\right)^2 - \frac{1}{\mu} \left(
    \frac{1-K(q)}{q^2}\right)^2 \rb\nn\\
&&\quad - 6 \frac{g^4}{N} \int_q \frac{1}{q^4} \lb \left(1 -
    \K{q}\right)^4 - \frac{1}{\mu} (1-K(q))^4 \rb\,.
\end{eqnarray}
\end{enumerate}

The asymptotic behaviors as $\Lambda \to \infty$ are given by
\begin{eqnarray}
    \frac{1}{\Lambda^2} u_2 (\Lambda; \bar{p} \Lambda,-\bar{p} \Lambda)
    &\longrightarrow& \frac{\lambda}{2} \frac{1}{\Lambda^2} \left[ 
        (\Lambda - \mu) \int_q \frac{\Delta (q)}{q^2} - \frac{m^2}{\mu} \int_q
        \frac{1 - K(q)}{q^4} \right] \nn\\
    && - 2 g^2 \frac{\Lambda - \mu}{\Lambda^2} \int_q \frac{\Delta (q)
      (1-K(q))}{q^2} \nn\\
    && + \frac{g^2}{\mu} \bar{p}^2 \left[ \frac{1}{2} \int_q
        \frac{(1-K(q))^2}{q^4} + c_2 \right]\,,\\ 
    z_2 (\Lambda; \bar{p}^2 \Lambda^2)
    &\longrightarrow& \frac{1}{N}
    \frac{g^2}{\mu} \left[ \int_q \frac{(1-K(q))^2}{q^4} + k_2 \right]\,,\\
    \frac{1}{\sqrt{\Lambda}} G(\Lambda; 0,0) &\longrightarrow& -
    \frac{g}{\sqrt{\Lambda}} - \frac{1}{N}
    \frac{g^3}{\mu \sqrt{\Lambda}} 
    \int_q \frac{(1-K(q))^3}{q^4}\,,\\
    \frac{1}{\Lambda} u_4 (\Lambda; 0,0,0,0) &\longrightarrow& -
    \frac{\lambda}{\Lambda} - \frac{3}{2} \frac{\lambda^2}{\mu \Lambda}
    \int_q \frac{(1-K(q))^2}{q^4} \nn\\
&& \quad +  \frac{6}{N} \frac{g^4}{\mu \Lambda} \int_q
    \frac{(1-K(q))^4}{q^4} \,.
\end{eqnarray}

\section{Analytic calculations of $m_{cr}^2 (\lambda, g^2)$ and
  $m_{tr}^2 (\lambda, g^2)$\label{app-transitions}}

The 1-loop RG equations can be solved analytically.  Hence, given
arbitrary $(\lambda, g^2)$ in Region 1, we can compute the critical
value $m_{cr}^2$.  Similarly, given $(\lambda, g^2)$ in Region 2 near
the line GB, we can compute the transition value $m_{tr}^2$.  In this
appendix, we first solve the flow equations, and then calculate
$m_{cr}^2$ and $m_{tr}^2$.

\subsection{Solving the flow equations}

The flow equations derived in \S \ref{RG equations} are given by
\begin{equation}
\lb\begin{array}{c@{~=~}l}
\frac{d}{dt} g^2 (t) & g^2 (t) \left( 1 - \frac{g^2
      (t)}{g_*^2}\right)\,,\\
\frac{d}{dt} \lambda (t) & \frac{1}{\lambda_I} \lb \lambda_+
    (g^2(t)) - \lambda (t) \rb \lb \lambda (t) - \lambda_- (g^2
    (t)) \rb\,,
\end{array}\right.
\end{equation}
where $\lambda_\pm (g^2)$ ($\lambda_+ > \lambda_-$) are defined by
\begin{equation}
\lb\begin{array}{c@{~=~}l}
 \lambda_+ (g^2) + \lambda_- (g^2) &
\lambda_I \lb 1 - g^2 \frac{1}{3} \left( I_3 + 2 I_5 \right)\rb\,,\\
\lambda_+ (g^2) \lambda_- (g^2) & - \lambda_I g^4 \frac{6}{N} I_7\,.
\end{array}\right.
\end{equation}

The solution for $g^2 (t)$ is given by
\begin{equation}
x (t) \equiv g^2 (t) \e^{-t} = \frac{g_*^2}{1 + \left(\frac{g_*^2}{g^2
        (0)} - 1 \right)\e^t}\,.
\end{equation}
We choose $0 < g^2 (0) < g_*^2$ so that 
\begin{equation}
x (t) \stackrel{t \to + \infty}{\longrightarrow} 0\,.
\end{equation}
Defining
\begin{equation}
f(t) \equiv \frac{\lambda (t)}{g^2 (t)}\,,
\end{equation}
we obtain
\begin{equation}
\frac{\frac{\lambda_{*+}}{g_*^2} - f(t)}{f(t) -
  \frac{\lambda_{*-}}{g_*^2}} = k \cdot
x(t)^{y_{\lambda -}}\,,
\label{app-fx}
\end{equation}
where $k$ is a constant determined by the initial condition, and
\begin{equation}
y_{\lambda-} \equiv \frac{\lambda_{*+}-\lambda_{*-}}{\lambda_I} > 0\,.
\end{equation}
\begin{figure}[t]
\begin{center}
\includegraphics[scale=1.2]{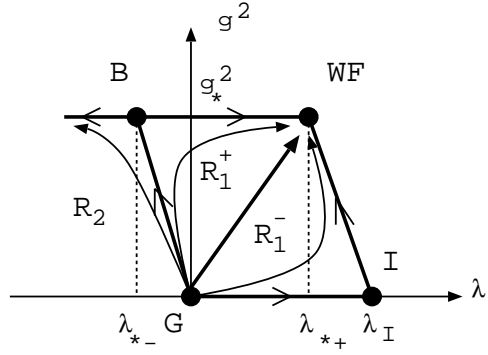}
\end{center}
\caption{\label{RGflows-app}RG flows}
\end{figure}
In Fig.~\ref{RGflows-app}, the three regions $R_1^+, R_1^-, R_2$ are
defined by the behavior of $f$:
\begin{equation}
\lb
\begin{array}{c@{\quad}l@{\quad}l}
R^-_1:& \frac{\lambda_{*+}}{g_*^2} < f,& f \stackrel{x\to
  0}{\longrightarrow} \frac{\lambda_{*+}}{g_*^2} + 0\,,\\
R_1^+:& \frac{\lambda_{*-}}{g_*^2} < f < \frac{\lambda_{*+}}{g_*^2},&
f \stackrel{x\to 0}{\longrightarrow} \frac{\lambda_{*+}}{g_*^2} - 0\,,\\
R_2:& f < \frac{\lambda_{*-}}{g_*^2}\,.
\end{array}\right.
\end{equation}
$k$ is positive in $R_1^+$, but negative in $R^-_1$ and $R_2$.  In $R_2$,
the region of a first order transition, we find $f \to - \infty$
(hence $\lambda \to - \infty$) at a finite $t$.

The flow equation for the squared mass is given by
\begin{equation}
\frac{d}{dt} m^2 (t) = \lb 2 +  A g^2 (t) + B \lambda (t)
 \rb m^2 (t) + C g^2 (t) + D \lambda (t)\,,
\end{equation}
where
\begin{equation}
A \equiv - \frac{1}{6} \left( I_3 + 2 I_5 \right),\quad
B \equiv - \frac{1}{2} I_4,\quad
C \equiv - 2 I_2,\quad
D \equiv \frac{1}{2} I_1\,.
\end{equation}
This is solved as
\begin{eqnarray}
&& \exp \left[ - 2 t - \int_0^t dt'\, \left( A g^2 (t') + B \lambda
        (t') \right) \right] m^2 (t)
 = m^2 (0) \nn\\
&&\quad + \int_0^t dt' \left( C g^2 (t') + D \lambda (t') \right)
\exp \left[ - 2 t' - \int_0^{t'} dt'' \left( A g^2 (t'') + B \lambda
        (t'') \right) \right].\label{m2-analytic}
\end{eqnarray}

\subsection{Critical squared mass $m_{cr}^2 ( \lambda, g^2)$}

In region $R_1 \equiv R_1^+ \cup R_1^-$, the critical value of $m^2 (0) = m^2$ is
obtained as a function of $g^2 (0) = g^2$ and $\lambda (0) = \lambda$
by the condition
\begin{equation}
m^2 (t) \stackrel{t \to +\infty}{\longrightarrow} m^2_{*+}\,.
\end{equation}
This implies that the left-hand side of (\ref{m2-analytic}) vanishes
in the limit $t \to +\infty$.  Hence, we obtain
\begin{equation}
m_{cr}^2 = - \int_0^\infty dt \left( C g^2 (t) + D \lambda (t) \right)
\exp \left[ - 2 t - \int_0^t dt' \left( A g^2 (t') + B \lambda (t')
    \right) \right]\,.
\end{equation}
Using the analytic solution for $g^2 (t)$ and $\lambda (t)$, this can
be rewritten as
\begin{equation}
m_{cr}^2 (\lambda, g^2) = - g_*^2 \int_0^1 \frac{dx}{x} \frac{C + D f(x)}{
  \left( 1 + \frac{g_*^2}{g^2} \left(\frac{1}{x} - 1 \right)
  \right)^2} \exp \left[ - g_*^2 \int_x^1 \frac{dx'}{x'} \left( A +
        B f(x') \right) \right]\,,
\end{equation}
where $f(x)$ is the same $f$ as before except it is regarded now as a
function of $x = g^2 (t)/g^2$, and it is explicitly given by
\begin{equation}
f(x) \equiv \frac{\left(\frac{\lambda}{g^2} -
      \frac{\lambda_{*-}}{g_*^2} \right) \cdot \frac{\lambda_{*+}}{g_*^2} +
  \left(\frac{\lambda_{*+}}{g_*^2} - \frac{\lambda}{g^2}\right)
   x^{y_{\lambda-}} \cdot \frac{\lambda_{*-}}{g_*^2}}{
\left(\frac{\lambda}{g^2} -
      \frac{\lambda_{*-}}{g_*^2}\right) + \left(\frac{\lambda_{*+}}{g_*^2} -
          \frac{\lambda}{g^2}\right)
      x^{y_{\lambda -}}} \,.
\end{equation}

\subsection{First order transition point $m_{tr}^2 (\lambda, g^2)$}

We consider a region directly below the trajectory GB, which is a line
given by
\begin{equation}
\frac{g^2}{g_*^2} = \frac{\lambda}{\lambda_{*-}}\,.
\end{equation}
For $g^2 < g_*^2$ and a positive infinitesimal $\ep$,
\begin{equation}
\lambda = \frac{\lambda_{*-}}{g_*^2} \left( g^2 +
        \ep \right) <  \frac{\lambda_{*-}}{g_*^2} g^2
\end{equation}
gives a point in $R_2$ just below the trajectory GB.  The squared mass
at the first order transition is obtained as
\begin{equation}
m_{tr}^2 \left(\frac{\lambda_{*-}}{g_*^2} \left( g^2 +
        \ep \right), g^2\right) \simeq m_{cr}^2 \left(
        \frac{\lambda_{*-}}{g_*^2} g^2, g^2 \right) +
    \frac{\lambda_{*-}}{g_*^2} \ep \frac{\partial}{\partial \lambda}
    m_{cr}^2 (\lambda, g^2) \Big|_{\lambda =
      \frac{\lambda_{*-}}{g_*^2} g^2 + 0}\,,
\end{equation}
where
\begin{eqnarray}
m_{cr}^2  \left(\frac{\lambda_{*-}}{g_*^2} g^2 , g^2\right)
&=& y_{E-} m_{*-}^2 \frac{\left(\frac{g^2}{g_*^2}\right)^2}{\left(1 -
      \frac{g^2}{g_*^2}\right)^2} \int_0^1 dx\,
\frac{x^{y_{E-}-1}}{\left(x - \frac{g_*^2}{g_*^2 - g^2}
  \right)^2}\nn\\
&=& y_{E-} m_{*-}^2 \frac{\left(\frac{g^2}{g_*^2}\right)^2}{\left(1 -
      \frac{g^2}{g_*^2}\right)^{y_{E-}}} \int_{\frac{g^2}{g_*^2}}^1
dx\, \frac{(1-x)^{y_{E-}-1}}{x^2}\,,
\end{eqnarray}
and the derivative is given by
\begin{eqnarray}
&&\frac{\partial}{\partial \lambda}
    m_{cr}^2 (\lambda, g^2) \Big|_{\lambda =
      \frac{\lambda_{*-}}{g_*^2} g^2 + 0} = \frac{B}{y_{\lambda-}}
    \frac{g_*^2}{g^2} m_{cr}^2 \left( g^2
        \frac{\lambda_{*-}}{g_*^2}, g^2 \right)\nn\\
&&\quad - \left(\frac{y_{E-}}{y_{\lambda-}} B m_{*-}^2 + D \right)
\frac{\frac{g^2}{g_*^2}}{\left(1 - \frac{g^2}{g_*^2}\right)^{y_{E-} -
    y_{\lambda-}}} \int_{\frac{g^2}{g_*^2}}^1 dx\,
\frac{(1-x)^{y_{E-}-y_{\lambda-}-1}}{x^2} \,.
\end{eqnarray}

\section{ERG for the $\mathcal{N} = 1$ Wess-Zumino model in $D=3$\label{app-WZ}}

To study the three dimensional $\mathcal{N} = 1$ Wess-Zumino model, it
is the most convenient if we preserve supersymmetry manifestly by
linearizing it with the help of a real auxiliary field $F$. The
corresponding classical lagrangian is
\begin{equation}
\mathcal{L}_{WZ} = \frac{1}{2} \nabla \phi \cdot \nabla \phi +
\frac{1}{2} \tilde{\chi} \vec{\sigma} \cdot \nabla \chi + \frac{1}{2} F^2
 + g \lb i F \frac{1}{2} \left( \phi^2 - v^2 \right) + \phi
\frac{1}{2} \tilde{\chi} \chi \rb\,.
\end{equation}
Integrating over $F$, we reproduce the classical action given by
(\ref{Wess-Zumino}) in sect.~\ref{Yukawa model}.  The imaginary $i$ is
necessary for the Euclidean space.  The classical lagrangian is invariant
under the following $\mathcal{N} = 1$ supersymmetry transformation:
\begin{equation}
\lb\begin{array}{c@{~=~}l}
\delta \phi & \tilde{\chi} \xi\,,\\
\delta \chi & \left( \vec{\sigma} \cdot \nabla \phi - i F\right) \xi\,,\\
\delta iF & \nabla \tilde{\chi} \cdot \vec{\sigma} \xi\,,
\end{array}\right.
\end{equation}
where $\xi$ is an arbitrary constant spinor. Besides supersymmetry,
the theory is invariant under the $\mathbf{Z_2}$ transformation:
\begin{equation}
\lb\begin{array}{c@{~\longrightarrow~}l}
 \phi (x) & - \phi (-x)\,,\\
 \chi (x) & i \chi (-x)\,,\\
 F (x) & F (-x)\,.
\end{array}\right.
\end{equation}
We construct a Wilson action that is invariant under the linearized
supersymmetry and $\mathbf{Z_2}$ transformation.  (The $\mathbf{Z_2}$
invariance forbids the supersymmetric mass term $iF\phi + \frac{1}{2}
\tilde{\chi} \chi$.)

The Wilson action is split into the free and interaction parts:
\begin{equation}
\SL = - \frac{1}{2} \int_p \frac{1}{\K{p}} \left(
p^2 \phi (-p) \phi (p) + \tilde{\chi} (-p) \vec{\sigma} \cdot i
\vec{p} \chi (p) + F(-p) F(p) \right) + \SI\,,
\end{equation}
where $\SI$ satisfies the ERG differential equation
\begin{eqnarray}
- \Lambda \frac{\partial}{\partial \Lambda} \SI &=& \frac{1}{2} \int_p
\frac{\Delta (p/\Lambda)}{p^2} \left[
\frac{\delta \SI}{\delta \phi (-p)} \frac{\delta \SI}{\delta \phi (p)}
+ \frac{\delta^2 \SI}{\delta \phi (-p) \delta \phi (p)}\right.\nn\\
&& - \Tr  (-i) \vec{p} \cdot \vec{\sigma} \lb
\Ld{\tilde{\chi} (-p)} \SI \cdot \SI \Rd{\chi (p)} 
+ \Ld{\tilde{\chi} (-p)} \SI \Rd{\chi (p)}\rb\nn\\
&& \left. + p^2 \lb \frac{\delta \SI}{\delta F (-p)} \frac{\delta \SI}{\delta
  F(p)} + \frac{\delta^2 \SI}{\delta F(-p) \delta F(p)} \rb \right]\,.
\end{eqnarray}

For UV renormalizability, we impose the following asymptotic behavior:
\begin{eqnarray}
    \SI &\stackrel{\Lambda \to \infty}{\longrightarrow}&
    - \int d^3 x\, \left[
        z_{UV} \frac{1}{2} \left( (\nabla \phi )^2 +  \tilde{\chi}
        \vec{\sigma} \cdot \nabla \chi + F^2 \right) \right.\nn\\
&&\quad\left.  +  g_{UV} \frac{1}{2} \left( iF \phi^2 + \phi \tilde{\chi} \chi
        \right) - (g v^2)_{UV} \frac{1}{2} i F \right]\,,
\end{eqnarray}
where $z_{UV}$ and $g_{UV}$ are constants, and $(g v^2)_{UV}$ is
linear in $\Lambda$.  (Note that in the common language the
coefficient of $iF(0)$ is linearly divergent; supersymmetry is
consistent with this divergence.)  The parameters $g, v^2$ are
introduced through the expansion:
\begin{eqnarray}
S_{I,\Lambda} &=& - \int_p z_2 (\Lambda; p)  \frac{1}{2} \left(
p^2 \phi (-p) \phi (p) + \tilde{\chi} (-p) i \vec{\sigma} \cdot \vec{p}
\chi (p) + F(-p) F(p) \right)\nn\\
&& \, - \int_{p,k} G (\Lambda; p,q) \, \frac{1}{2} \left(
iF (-p-q) \phi (p) \phi (q) + \phi (-p-q) \tilde{\chi} (p) \chi (q)
\right)\nn\\
&&\, + (GV^2) (\Lambda) \frac{1}{2} i F(0)\,,
\end{eqnarray}
where we impose
\begin{equation}
\lb\begin{array}{r@{~=~}l}
z_2 (\mu; 0) & 0\,,\\
G(\mu; 0,0) & g\,,\\
(GV^2) (\mu) & g v^2\,,
\end{array}\right.
\end{equation}
at an arbitrary renormalization scale $\mu$.

The dependence of $S_\Lambda$ on $\mu$ is given by the RG equation
\begin{equation}
- \mu \frac{\partial}{\partial \mu} \SL = - \beta \partial_g \SL -
\beta_{v^2} \partial_{v^2} \SL + \gamma \N\,,
\end{equation}
where $\N$ is the equation-of-motion operator that counts the number of
fields.  The 1-loop calculations give the coefficients as
\begin{equation}
\begin{array}{c@{~=~}l}
\beta & - \frac{g^3}{\mu} \left( I_6 + \frac{3}{4} I_5 \right)\,,\\
\beta_{v^2}& - \mu I_1 + \frac{g^2}{\mu} v^2 \left(I_6 + \frac{1}{2}
  I_5\right)\,, \\
\gamma & \frac{1}{4} \frac{g^2}{\mu} I_5\,,
\end{array}
\end{equation}
where the integrals are defined in Appendix \ref{app-integrals}.
Unlike the $\mathcal{N} = 1$ Wess-Zumino model in $4$ dimensions or
its dimensionally reduced $\mathcal{N}=2$ model in $3$ dimensions,
there is no non-renormalization theorem for this model.

Redefining $g^2/\mu$ by $g^2$ and $v^2/\mu$ by $v^2$, the RG flows are given by
\begin{equation}
\begin{array}{c@{~=~}l}
\frac{d}{dt} g^2& g^2 + 2 g \beta\,,\\ 
\frac{d}{dt} v^2 & v^2 + \beta_{v^2}\,.
\end{array}
\end{equation}
Besides the Gaussian fixed point $g^2 = v^2 = 0$, the flows have a
non-trivial fixed point at
\begin{equation}
\begin{array}{c@{~\simeq~}l}
g_*^2 & \frac{1}{2} \frac{1}{I_6 + \frac{3}{4} I_5}\,,\\
v_*^2 & I_1\,.
\end{array}
\end{equation}
The anomalous dimension of the elementary fields is given by
\begin{equation}
\eta = 2 \gamma (g_*^2) = \frac{1}{4} \frac{I_5}{I_6 + \frac{3}{4}
  I_5}\,,
\end{equation}
and the scale dimension of $v^2 - v_*^2$ is
\begin{equation}
y_E = 1 + \frac{1}{2} \frac{I_6 + \frac{1}{2} I_5}{I_6 + \frac{3}{4}
  I_5}\,.
\end{equation}
These 1-loop results confirm the sum rule found in
Ref.~\citen{Synatschke:2010ub}:
\begin{equation}
2 y_E + \eta = 3
\end{equation}
which is valid for any choice of the cutoff function in our case.
Using the particular cutoff function given in Appendix
\ref{app-integrals}, we obtain the plot in Fig.~\ref{plot-yEWZ}.
\begin{figure}[t]
\begin{center}
\includegraphics[scale=0.6]{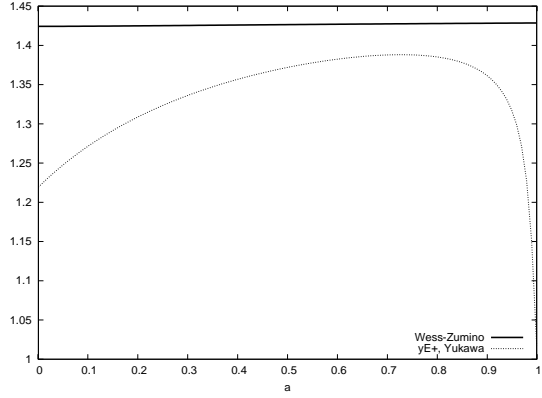}
\end{center}
\caption{\label{plot-yEWZ}The exponent $y_E \simeq 1.43$ for the
  Wess-Zumino model 
is  compared with $y_{E+} \simeq 1.38$ of the Yukawa model.}
\end{figure}
The exponent $y_E \simeq 1.43$ hardly depends on the cutoff parameter,
and it agrees a little better with $y_{E+} = 1.408$ from
Ref.~\citen{Synatschke:2010ub} than $y_{E+} \simeq 1.38$ of the $N=1$
Yukawa model.

\section{Integrals of a cutoff function\label{app-integrals}}

We define the following integrals:
\begin{equation}
\begin{array}{c@{~\equiv~}l@{\quad}c@{~\equiv~}l}
I_1 & \int_q \frac{\Delta (q)}{q^2}\,,& I_2 & \int_q \frac{\Delta
  (q)(1-K(q))}{q^2}\,,\\
I_3 & \int_q \frac{\Delta (q)^2}{q^4}\,,& I_4 & \int_q \frac{1-K
  (q)}{q^4}\,,\\
I_5 & \int_q \frac{(1-K(q))^2}{q^4}\,,& I_6 & \int_q
\frac{(1-K(q))^3}{q^4}\,,\\
I_7 & \int_q \frac{(1-K(q))^4}{q^4}\,.
\end{array}
\end{equation}

\begin{figure}
\begin{center}
\includegraphics[scale=1.2]{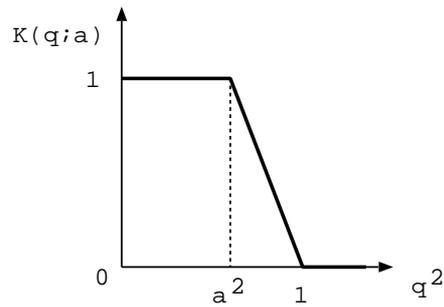}
\end{center}
\caption{\label{app-Ka}A cutoff function $K(q;a)$ parametrized by $0 <
  a < 1$}
\end{figure}

For a particular choice of the cutoff function (see Fig.~\ref{app-Ka})
\begin{equation}
K(q;a) \equiv \lb\begin{array}{c@{\quad\textrm{for}\quad}l}
1 & 0 < q^2 < a^2\,,\\
\frac{1-q^2}{1-a^2}& a^2 < q^2 < 1\,,\\
0 & 1 < q^2\,,
\end{array}\right.
\end{equation}
the above integrals can be evaluated easily as follows:
\begin{equation}
\begin{array}{c@{~=~}l@{\quad}c@{~=~}l}
I_1 (a) & \frac{1}{2 \pi^2} \cdot \frac{2}{3} \frac{1 + a + a^2}{1+a},&
I_2 (a) & \frac{1}{2 \pi^2} \cdot \frac{2}{15} \frac{2 a^3 + 4 a^2 + 6
  a + 3}{(1+a)^2},\\
I_3 (a) & \frac{1}{2 \pi^2} \cdot \frac{4}{3} \frac{1 + a +
  a^2}{(1+a)^2 (1-a)},&
I_4 (a) & \frac{1}{2 \pi^2} \cdot \frac{2}{1+a},\\
I_5 (a) & \frac{1}{2 \pi^2} \cdot \frac{4}{3} \frac{1+2 a}{(1+a)^2},&
I_6 (a) & \frac{1}{2 \pi^2} \cdot \frac{6 + 18 a + 16 a^2}{5(1+a)^3},\\
I_7 (a) & \frac{1}{2 \pi^2} \cdot \frac{8}{35} \frac{5 + 20 a + 29 a^2
  + 16 a^3}{(1+a)^4} \,.
\end{array}
\end{equation}

\end{document}